# Superconductivity in novel quasi-one-dimensional ternary molybdenum pnictides $Rb_2Mo_3As_3$ and $Cs_2Mo_3As_3$

Kang Zhao[1,2,†], Qing-Ge Mu[1,2,†], Tong Liu[1,2], Bo-Jin Pan[1,2], Bin-Bin Ruan[1,2], Lei Shan[1,2,3], Gen-Fu Chen[1,2,3], and Zhi-An Ren[1,2,3,*]

[1] Institute of Physics and Beijing National Laboratory for Condensed Matter Physics, Chinese Academy of Sciences, Beijing 100190, China

[2] School of Physical Sciences, University of Chinese Academy of Sciences, Beijing 100049, China

[3] Collaborative Innovation Center of Quantum Matter, Beijing 100190, China.

[†] *These authors contributed equally to this work.*

[*]Email: renzhian@iphy.ac.cn

Keywords: superconductivity, $Rb_2Mo_3As_3$, $Cs_2Mo_3As_3$, quasi-one-dimensional

**Abstract**

By replacing the alkali element in the newly discovered $K_2Mo_3As_3$ superconductor, we successfully synthesized ternary molybdenum pnictides $Rb_2Mo_3As_3$ and $Cs_2Mo_3As_3$ through solid state reaction method. Powder X-ray diffraction analysis reveals the same quasi-one-dimensional (Q1D) hexagonal crystal structure and space group of *P*-6*m*2 (No. 187) as $K_2Mo_3As_3$. The refined lattice parameters are $a$ = 10.432 (1) Å, $c$ = 4.4615 (6) Å for $Rb_2Mo_3As_3$ and $a$ = 10.7405 (6) Å, $c$ = 4.4654 (5) Å for $Cs_2Mo_3As_3$. Electrical resistivity and magnetic susceptibility characterizations exhibit the occurrence of superconductivity in both compounds with the onset $T_c$ at 10.6 K and 11.5 K for $Rb_2Mo_3As_3$ and $Cs_2Mo_3As_3$ respectively, which exhibit weak negative chemical pressure effect in these $A_2Mo_3As_3$ (A = K, Rb, Cs) superconductors contrary to the isostructural $A_2Cr_3As_3$ superconductors. More interestingly, the $Cs_2Mo_3As_3$ superconductor exhibits much higher upper critical field around 60 T at zero temperature. The discovery of these MoAs/CrAs-based superconductors provide a unique platform for the study of exotic superconductivity correlated with both 3d and 4d electrons in these Q1D compounds.

## 1. Introduction

As the high-$T_c$ superconductivity is found in cuprates and iron pnictides/chalcogenides which have typical layered quasi-two-dimensional crystal structure, transition metal based chemical compounds with low dimensionality are more and more considered to be the interesting research objects for the discovery of exotic superconductivity or other physical phenomenon [1-4]. Recently, a novel family of chromium arsenide based superconductors $A_2Cr_3As_3$ (A = Na, K, Rb or Cs) were discovered, which contains special quasi-one-dimensional (Q1D) $(Cr_3As_3)^{2-}$ linear chains separated by columns of alkali metal cations [5-8]. These Q1D superconductors exhibit some intriguing characteristics from their physical properties characterizations and possible unconventional spin-triplet superconductivity was suggested [9-16]. Lately, superconductivity was also reported in some Q1D-type $ACr_3As_3$ crystals which have identical crystal lattice to the Mo-based $M_2Mo_6X_6$ superconductors discovered in 1980's [17-20]. These Cr-233 and Cr-133 superconductors exhibit similar superconductivity with some different behaviors and need more investigations. Although the Mo-based ternary compounds have been thoroughly studied for the occurrence of superconductivity in many molybdenum chalcogenides (Chevrel phases) from 1970's [21-23], the first MoAs-based ternary superconductor $K_2Mo_3As_3$ was only discovered very recently [24]. After that, we studied the replacement of $K^+$ cation with other alkali elements, and successfully synthesized new ternary molybdenum pnictides $Rb_2Mo_3As_3$ and $Cs_2Mo_3As_3$ through solid state reaction method, with superconductivity observed in both compounds revealed by electrical resistivity and magnetic susceptibility measurements.

## 2. Experimental

Polycrystalline $Rb_2Mo_3As_3$ and $Cs_2Mo_3As_3$ samples were synthesized through high temperature solid state reaction using Rb (pieces, 99%), Cs (pieces, 99%), Mo (powder, 99.95%), and As (powder, 99.999%) elements as the starting materials. At first, the RbAs and CsAs were synthesized by mixing Rb or Cs pieces and As with a stoichiometric ratio of 1:1 and then sealed into an evacuated quartz tube. The quartz tube was very slowly heated to 523 K and sintered for 20 hours followed by furnace cooling down to room temperature. The obtained RbAs or CsAs was ground into fine powder and mixed with Mo and As powder with a stoichiometric ratio of 2.3:3:0.7 and pressed into a small pellet. The pellet was placed into an alumina crucible, and sealed into a Ta tube with arc welding in argon atmosphere. The sample was heated at 1123 K for 50 hours before cooled down to room temperature by furnace shut-down. After that, polycrystalline samples of $Rb_2Mo_3As_3$ and $Cs_2Mo_3As_3$ were obtained for structural and physical characterizations. All experimental procedures mentioned above except sealing and sintering were carried out in a glove box filled with high pure argon atmosphere to avoid the possible oxidation to the samples.

The obtained samples are extremely reactive in air, and hence any exposure to air should be avoided when performing measurements on these samples. The crystal structure was characterized at room temperature by powder x-ray diffraction (XRD) using a PAN-analytical x-ray diffractometer with Cu-$K_\alpha$ radiation. The sample morphology was characterized with a Phenom scanning electron microscope (SEM). The electrical resistivity was measured on a Quantum Design physical property measurement system with the standard four-probe method under a magnetic field from 0 T to 9 T. The dc magnetization was measured with a Quantum Design magnetic property measurement system under zero-field-cooling (ZFC) and field-cooling (FC) modes.

3. Results and discussion

The as-grown samples of both $Rb_2Mo_3As_3$ and $Cs_2Mo_3As_3$ compounds are black in color with good ductility. In Fig. 1 we show the typical powder XRD patterns collected from 8° to 90° at room temperature. The observed diffraction peaks were well indexed with the refined Bragg lattice calculated from the space group *P-6m2* (No. 187), indicating the same hexagonal structure with $K_2Mo_3As_3$. Minor impurity phase of arsenic can be occasionally observed. The inset SEM images in Fig. 1 show the fresh fracture surfaces for both samples, which exhibit all needle-like crystal grains and further evidence the Q1D structure in which $(Mo_3As_3)^{2-}$ linear chains are separated by $Rb^+$ or $Cs^+$ cations. The refined lattice parameters are $a$ = 10.432 (1) Å, $c$ = 4.4615 (6) Å for $Rb_2Mo_3As_3$ and $a$ = 10.7405 (6) Å, $c$ = 4.4654 (5) Å for $Cs_2Mo_3As_3$. For these novel $A_2Mo_3As_3$ (A = K, Rb, Cs) compounds, the crystal structure expands clearly along the *a-axis* while the *c-axis* changes very slightly by the replacement of alkali cations with larger radius, which is similar to that of the $A_2Cr_3As_3$ compounds since the inter-chain bonding is dominated by the alkali cations. The excess using of alkali metals in the starting materials is important to compensate the volatilization and to obtain single-phase samples.

The typical temperature dependences of electrical resistivity for $Rb_2Mo_3As_3$ and $Cs_2Mo_3As_3$ are displayed in Fig. 2, and both compounds exhibit clear superconducting transitions at low temperature. The onset superconducting $T_c$ for $Rb_2Mo_3As_3$ is 10.6 K, for $Cs_2Mo_3As_3$ it is 11.5 K, and the $Rb_2Mo_3As_3$ has a sharp transition of 0.4 K. The $T_c$ increases slightly by the replacement of alkali metals with larger ionic radius in $A_2Mo_3As_3$, which exhibit a negative chemical pressure effect contrary to the isostructural $A_2Cr_3As_3$ superconductors, but similar to the 133-type $ACr_3As_3$ superconductors. The different behaviors of the $T_c$ on the tuning of crystal structure may reflect the complicated electronic structure at Fermi energy in these Cr/Mo based Q1D compounds. The resistivity curves at normal state above $T_c$ are shown as the insets. We note that due to the very high activity of these samples, the normal state resistivity measurements have very poor reproducibility, and the measurements may not reflect the intrinsic nature of the sample resistivity. To characterize the upper critical field $\mu_0H_{c2}$, the electrical resistivity below 12 K was measured with temperature sweeping for both samples under constant magnetic fields with the direction perpendicular to the electrical current. The fields were varied from 0 T to 9 T with 1 T interval, and the data are shown in Fig. 3. For both samples, the magnetic field suppresses the superconducting transitions gradually, with an obvious broadening effect similar to $K_2Mo_3As_3$ and $Tl_2Mo_6Se_6$. It is worth to note that For $Cs_2Mo_3As_3$ a second transition can be observed under higher fields. No obvious magnetoresistance effect appears at the normal state. The data of $\mu_0H_{c2}$ vs. $T$ are both fitted using the formula from Ginzburg-Landau theory, $\mu_0H_{c2}(T) = \mu_0H_{c2}(0)(1-t^2)/(1+t^2)$, where $t$ stands for $T/T_c$. We concluded the $\mu_0H_{c2}(0)$ to be 23.4 T for $Rb_2Mo_3As_3$ and 60.2 T for $Cs_2Mo_3As_3$ by the G-L fit. Surprisingly, the upper critical field at zero temperature for $Cs_2Mo_3As_3$ is interestingly higher than that of $K_2Mo_3As_3$ and $Rb_2Mo_3As_3$, and also far above the Pauli paramagnetic limited critical field $\mu_0H_p = 1.84T_c = 21.2$ T. The behavior of the upper critical field of these samples indicates possible unconventional superconductivity in these Mo-233 type novel superconductors.

The temperature dependences of dc susceptibility from 2 K to 15 K for $Rb_2Mo_3As_3$ and $Cs_2Mo_3As_3$ with both ZFC and FC measurements under stable magnetic fields are displayed in Fig. 4. The data show clear diamagnetic superconducting transitions at 10.5 K and 11.5 K for $Rb_2Mo_3As_3$ and $Cs_2Mo_3As_3$ respectively. The results agree well with the electrical resistivity measurements of the superconducting $T_c$. Compared with the previous $K_2Mo_3As_3$ superconductor, both samples show much lower diamagnetic shielding volume fraction derived from the ZFC data. Especially for $Cs_2Mo_3As_3$, both the resistivity and susceptibility data indicate filamentary superconductivity in the sample. As reported in the $Tl_2Mo_6Se_6$ type superconductors, this is possibly due to the deviation of chemical composition from ideal stoichiometry for these samples which is very sensitive for superconductivity as observed in our experiments, it need to be further clarified by the growth of high quality samples.

In summary, we successfully synthesized new MoAs-based ternary compounds $Rb_2Mo_3As_3$ and $Cs_2Mo_3As_3$ which have the hexagonal Q1D crystal structure by solid state reaction method. Electrical resistivity and magnetic susceptibility characterizations revealed the occurrence of superconductivity with the onset $T_c$ at 10.6 K and 11.5 K for $Rb_2Mo_3As_3$ and $Cs_2Mo_3As_3$ respectively. The critical temperature in these $A_2Mo_3As_3$ (A = K, Rb, Cs) superconductors exhibit weak negative chemical pressure effect contrary to the previously reported isostructural $A_2Cr_3As_3$ superconductors. Among these MoAs-based novel superconductors, the $Cs_2Mo_3As_3$ exhibits interestingly higher upper critical field around 60 T. Containing the same group VIB transition elements, the discovery of these MoAs-based and previously CrAs-based superconductors provide the unique platform for the study of exotic superconductivity correlated with 3d and 4d electrons in these Q1D compounds.

**Acknowledgements**

The authors are grateful for the financial supports from the National Natural Science Foundation of China (No. 11474339 and 11774402), the National Basic Research Program of China (973 Program, No. 2016YFA0300301) and the Youth Innovation Promotion Association of the Chinese Academy of Sciences.

**Figure 1**

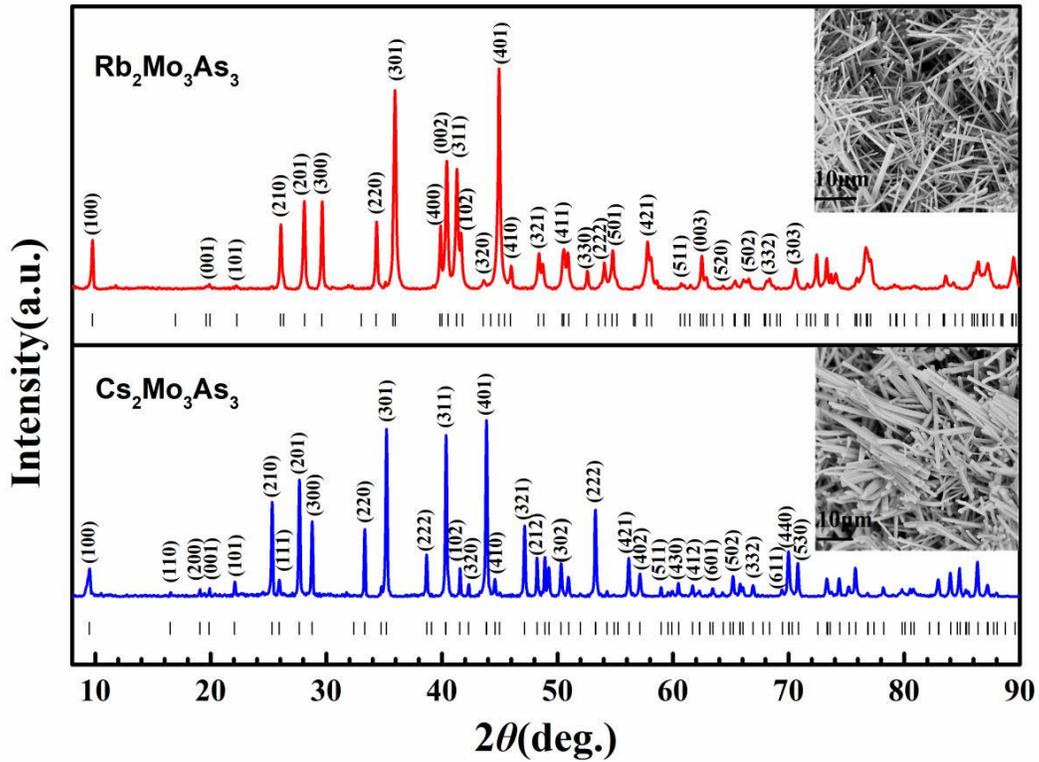

**Figure 2**

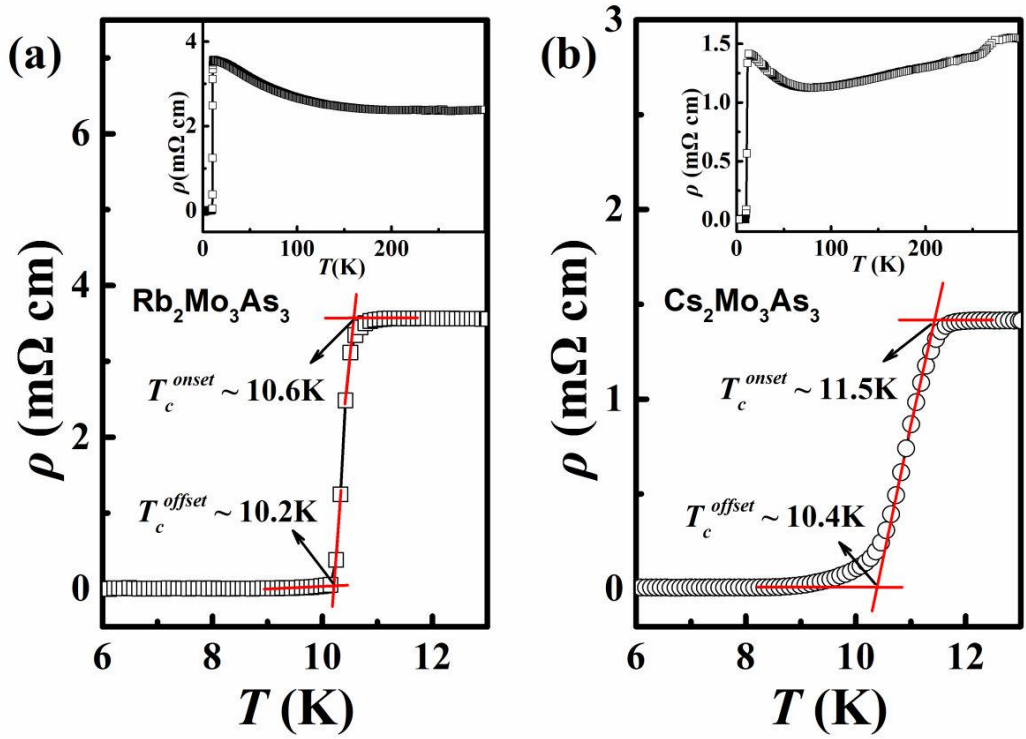

**Figure 3**

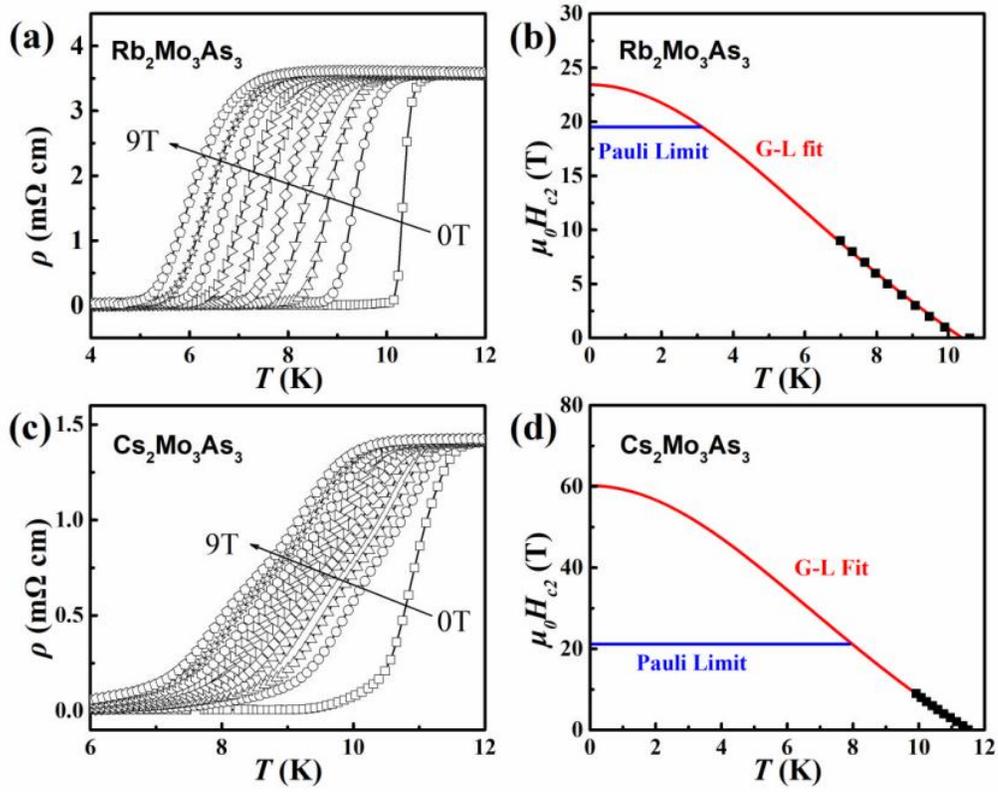

**Figure 4**

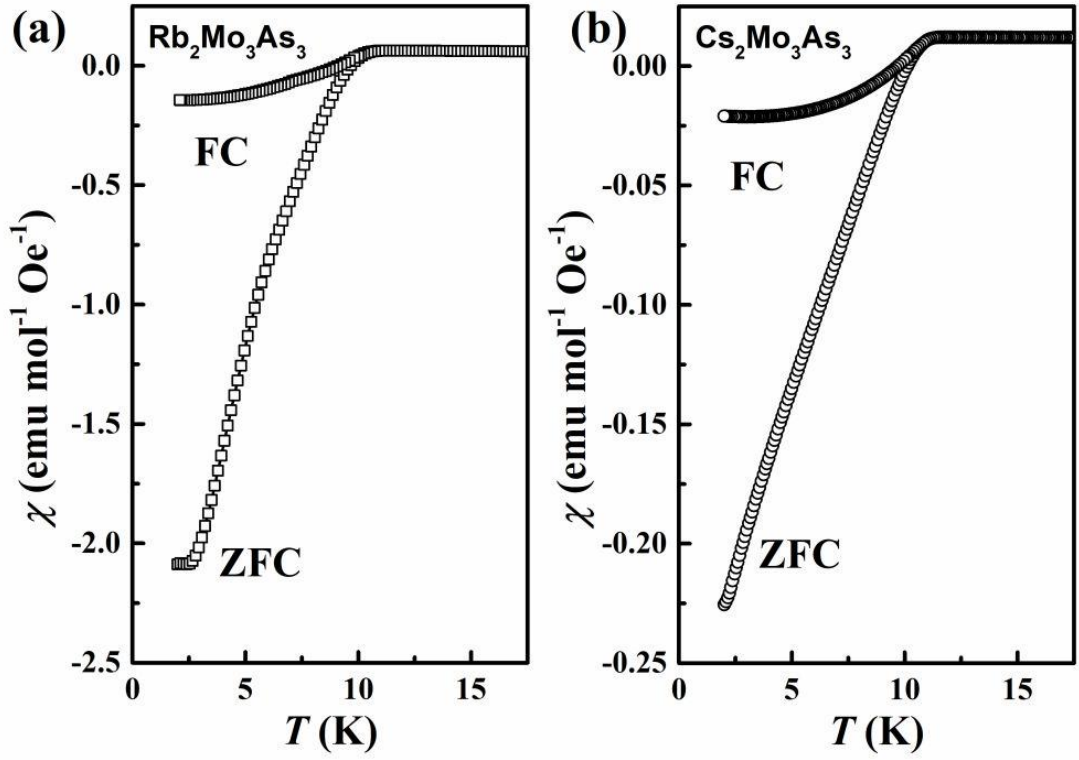